\documentclass[prb,nofootinbib,twocolumn,superscriptaddress]{revtex4} 

\usepackage{graphicx}
\usepackage{dcolumn}
\usepackage{bm}
\usepackage{threeparttable}
\usepackage{times}
\usepackage{mathptmx}
\usepackage{lscape}
\usepackage{natbib}
\usepackage{amsmath}
\usepackage{amssymb}
\usepackage{braket}
\usepackage{comment}

\def\degree{\kern-.2em\r{}\kern-.3em}

\begin{document}


\title{ Short-Range-Order for fcc-based binary alloys Revisited from Microscopic Geometry  }

\author{Koretaka Yuge}
\affiliation{
Department of Materials Science and Engineering,  Kyoto University, Sakyo, Kyoto 606-8501, Japan\\
}%

\begin{abstract}
{ Short-range order (SRO) in disordered alloys is typically interpreted as competition between chemical effect of negative (or positive) energy gain by mixing constituent elements and geometric effects comes from difference in effective atomic radius. Although we have a number of theoretical approaches to quantitatively estimate SRO at given temperatures, it is still unclear to systematically understand trends in SRO for binary alloys in terms of geometric character, e.g., effective atomic radius for constituents. Since chemical effect plays significant role on SRO, it has been believed that purely geometric character cannot quantitatively explain the SRO trends. Despite these considerations, based on the density functional theory (DFT) calculations on fcc-based 28 equiatomic binary alloys, we find that while convensional Goldschmidt or DFT-based atomic radius for constituents have no significant correlation with SRO, atomic radius for \textit{specially selected} structure, constructed purely from information about underlying lattice, can successfully capture the magnitude of SRO. These facts strongly indicate that purely geometric information of the system plays central role to determine characteristic disordered structure.  }
\end{abstract}


\maketitle

\section{Introduction}
In earlier days, they have attempted to systematically understand the tendency of SRO by ratio of atomic radius for disordered binary alloys, where it is expected that large difference in atomic radius promotes ordering tendency due mainly to reducing strain energy by forming neighboring unlike-atom pairs. 
However, these attempts have not captured even whether constituents are likely to exhibit ordering or clustering tendency, indicating that another effects of chemical ordering play central role for many binary alloys.
The importance of chemical effects, including many-body interactions in the system, on SRO has been pointed out by a variety of theoretical studies. 
Thus, so far, the geometric effects for binary alloys have not get so much attention to systematically understand the trends in SRO. 

Very recently, we have revealed that temperature dependence of SRO can be well-characterized by a single, special microscopic state (called projection state, PS) whose structure can be known \textit{a priori} without any information about energy or tempearture.
This is derived by clarifying how spatial constraint on the system connects with structure in equilibrium state, which provides new insight into SRO based on the information of configurational density of states for non-interactinc system.
These facts strongly imply that geometric effects on trends in SRO for binary alloys should be carefully re-examined, since the spatial constraint directly connects with the underlying geometric nature. 
In the present study, based on the theoretical approach we have developed combined with density functional theory (DFT) calculation, we first estimate magnitude relationship of SRO for fcc-based 28 equiatomic binary alloys. We demonstrate that trends in SRO cannot be explained by convensional Goldschmidt atomic radius or by DFT-based atomic radius even qualitatively, which is consistent with the previously-known results. Despite this fact, we find that trends in SRO have significant linear correlation with effective atomic radius for specially selected miscroscopic structure, which reveals that SRO for binary alloys can be well characterized by purely geometric character of the system. The details are given below. 

\section{Methodology}
We first consider a complete set of coordination, $\left\{q_{1},\ldots,q_{g}\right\}$, to completely describe possible microscopic structures on the system. 
Our previous study reveals that canonical average of SRO along chosen coordination $r$ can be given by
\begin{eqnarray}
\label{eq:lsi}
Q_{r}\left(T\right) \simeq \Braket{q_{r}}_{1} \mp \sqrt{\frac{\pi}{2}}\Braket{q_{r}}_{2} \cdot \beta \left( U_r^{\left( \pm \right)} - U_0 \right) 
\end{eqnarray}
where $\Braket{\quad}_1$ and $\Braket{\quad}_2$ denotes taking arithmetic average and standard deviation over all possible microscopic states on configuration space. 
$U_{r}$ and $U_0$ represents potential energy of PS and special quasirandom structure (SQS) that mimic perfect random state. 
Since PS, SQS, $\Braket{q_{r}}_1$, $\Braket{q_{r}}_2$ can be determined through configurational density of states for non-interacting system, we can \textit{a priori} know their value without any information about interactions or temperature. 
Explicitly, structure of PS and SQS is respectively given by $\left\{ \Braket{q_{1}}_{1}^{\left( r\pm \right)}, \cdots, \Braket{q_{f}}_{1}^{\left( r\pm \right)} \right\}$ and $\Braket{q_1}_1,\cdots, \Braket{q_f}_1$, where $\Braket{\cdot}_1^{\left( r+ \right)}$ ($\Braket{\cdot}_1^{\left( r- \right)}$ )denote taking linear average over possible microscopic sturctures that satisfy $q_r\ge \Braket{q_r}_1$ ($q_r\le \Braket{q_r}_1$).  

In the present study, we consider fcc-based equiatomic binary alloys including all possible combination of constituent elements of Ag, Al, Au, Cu, Ir, Pd, Pt and Rh, resulting in 28 binary system.
To quantitatively describe atomic configurations on fcc lattice, we employ generalized Ising model (GIM), providing a set of complete orthonormal basis functions: In A-B binary system, occupation of a lattice site $i$ is specified by spin variables, $\sigma_{i}=+1$ for A and $\sigma_{i}=-1$ for B. 
Using the spin variables, basis functions, 
$\phi_{s}^{\left(d\right)}$ is given by
\begin{eqnarray}
\label{eq:phi}
\phi_{s}^{\left(d\right)} = \left<\prod_{i\in s}\sigma_{i}\right>_{d}. 
\end{eqnarray}
Here, $\left<\quad\right>_{d}$ means taking average for microscopic state $d$, summations is taken over symmetry-nonequivalent figure $s$ consisting of multiple lattice points. 
Here, the structure of PS and SQS on fcc lattice along 1st nearest-neighbor (1NN) coordination is numerically constructed based on Monte Carlo simulation, to minimize difference between ideal and simulated value of $q_r$s, where we consider up to 6NN pair, and all triples and quartets consisting of up to 4NN pairs. 
Note that from Eq.~(\ref{eq:lsi}), we should construct two types of PS individually having $\left\{ \Braket{q_{1}}_{1}^{\left( r+ \right)}, \cdots, \Braket{q_{f}}_{1}^{\left( r+ \right)} \right\}$ and 
$\left\{ \Braket{q_{1}}_{1}^{\left( r- \right)}, \cdots, \Braket{q_{f}}_{1}^{\left( r- \right)} \right\}$. 
The constructed PS and SQS are then used for DFT calculation to obtain total energy for the 28 binary alloys, which is applied to Eq.~(\ref{eq:lsi}) to determine temperature-dependent SRO. 

In the DFT calculation, total energy is  estimated by the VASP\cite{vasp} code using the projector-augmented wave method,\cite{paw} with the exchange-correlation functional treated within the generalized-gradient approximation of Perdew-Burke-Ernzerhof (GGA-PBE).\cite{pbe}  The plane wave cutoff of 360 eV is used. 
In order to see the geometric effects on SRO in terms of individual contribution from changes in volume and that in internal atomic positions, we perform two types of DFT calculation: One is full structural optimization with the residual forces less than 0.001 eV/angstrom, and another is volume (and shape of the cell) optimization where internal atomic posistions are kept fixed at ideal lattice points.


\section{Results and Discussions}
\begin{figure}[h]
\begin{center}
\includegraphics[width=0.7\linewidth]
{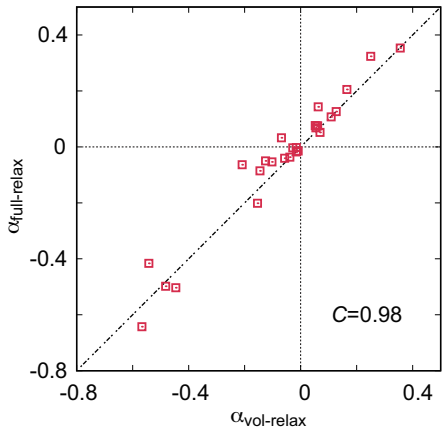}
\caption{Relationship between $\alpha$ obtained by volume-relaxed DFT calculation and that by full-relax DFT calculation in Fig.~\ref{fig:vf} for 28 fcc-based binary alloys. Correlation coefficient for the distribution, $C$, is given together.}
\label{fig:vf}
\end{center}
\end{figure}
\begin{figure}[h]
\begin{center}
\includegraphics[width=0.55\linewidth]
{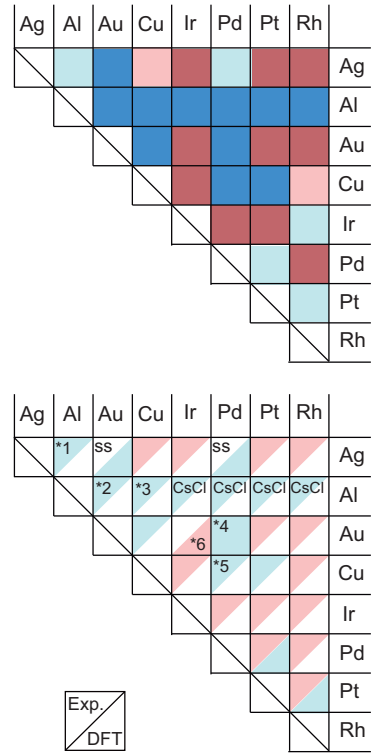}
\caption{Color plot of $\alpha$ (blue denotes $\alpha<0$, i.e., unlike-atom pair ordering, while red denotes $\alpha >0$, i.e., like-atom pair ordering (clustering)) obtained by the present approach (upper figure) and by the previous experimental\cite{pe1,pe2} and/or DFT studies\cite{pd1} (lower figure). 
\\${}^{*1}$Ag$_{0.67}$Al$_{0.33}$ (hcp) + Al (fcc). ${}^{*2}$ AuAl (orthorhombic). ${}^{*3}$ AlCu (monoclinic, mS20). ${}^{*4}$ AuPd (unknown prototype). 
${}^{*5}$ CuPd (CsCl type, B2). ${}^{*6}$ Replusive neighboring interaction predicted by the empirical, modified embedded atom method.\cite{pd2} 
}
\label{fig:table}
\end{center}
\end{figure}
We first see whether effect of changes in volume or that in internal atomic positions is dominant to capture the characteristics of SRO for the 28 fcc-based binary alloys. 
From Eq.~(\ref{eq:lsi}), it is clear that at given temperature, relative magnitude of SRO is completely specified by difference of energy between PS and SQS, $\alpha$, defined as
\begin{eqnarray}
\alpha= \left\{
\begin{array}{ll}
U_r^+ - U_0\quad \left( U_r^+ < U_r^- \right)  \\
-\left(U_r^- - U_0\right)\quad \rm{otherwise}
\end{array} \right.
\end{eqnarray}
With this definition, negative sign of $\alpha$ corresponds to preference of unlike-atom pair, and positive sign to like-atom pair. 
We therefore show relationship between $\alpha$ obtained by volume-relaxed DFT calculation and that by full-relax DFT calculation in Fig.~\ref{fig:vf}.
We can clearly see that $\alpha_{\textrm{vol-relax}}$ has significant linear correlation ($C=0.98$) with $\alpha_{\textrm{full-relax}}$, indicating that geometric effects on trends in SRO can be reasonablly caputured mainly by effects of changes in volume. 

In order to further see how trends in SRO connets with stables phases, we show in Fig.~\ref{fig:table} color plot of $\alpha$ (blue denotes $\alpha<0$, i.e., unlike-atom pair ordering, while red denotes $\alpha >0$, i.e., clustering) obtained by the present approach (upper figure) and by the previous experimental and/or DFT studies (lower figure). Dark blue (dark red) squares correspond to larger magnitude of $\alpha$ than light blue (light red) ones.
We can see that except for Rh-Ir binary alloy, sign of $\alpha$ has clear correlation with stable phases predicted by reported experiment and/or DFT calculations: Binarys alloy with negative value of $\alpha$ have stable ordered structure at equiatomic composition, while those with positive tend to undergo into phase separation of constituents. 
Since whether the system exhibit ordered structure or phase separation typically has strong correlation with the sign of SRO parameter,\cite{zun1,zun2} our predicted SRO can reasonablly capture the overall trends of SRO for the selected fcc-based alloys. 

Based on the results of SRO, we next see the relationship between $\alpha$ and ratio of atomic radius for constituents A and B, $R$. 
Here, $R=R_{\textrm{A}}/R_{\textrm{B}}$ ($R=R_{\textrm{B}}/R_{\textrm{A}}$) when $R_{\textrm{A}} > R_{\textrm{B}}$ ($R_{\textrm{B}} > R_{\textrm{A}}$). 
We consider two types of atomic radius for constituents, one is the convensional Goldschmidt atomic radius, and another is those obtained by DFT calculation.
Since the constituent elements of Ag, Al, Au, Cu, Ir, Pd, Pt and Rh all takes fcc, in DFT calculation, $R$ is determined by taking cube root of volume of the used cell. 
\begin{figure}[h]
\begin{center}
\includegraphics[width=0.85\linewidth]
{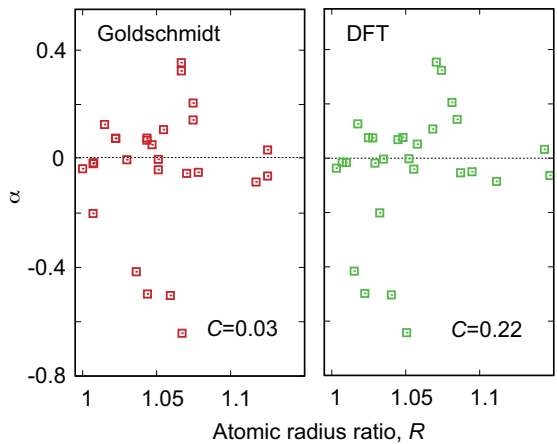}
\caption{Correlation between ratio of atomic radius (left: Goldschmidt, right: DFT calculation) for constituents and $\alpha$ for 28 binary alloys. Linear correlation coefficient, $C$, is given together.  }
\label{fig:rgd}
\end{center}
\end{figure}
The results are shown in Fig.~\ref{fig:rgd}. We can clearly see that no significant linear correlation between atomic radius ratio and SRO is found for Goldschmidt or DFT calclation. It has been considered that large difference of atomic radius (i.e., large value of $R$) leads to ordering tendency (corresponding to $\alpha < 0$) to effectively reduce strain energy due to size mismatch of constituents.
Figure~\ref{fig:rgd} cannot capture such tendency even qualitatively: Binary alloys with larger $R$ both have $\alpha$ in positive as well as in negative sign. 
These strongly indicate that atomic radius of constituents defined on unary system is not appropriate to explain the difference of SRO for binary alloys, which is consistent with the statement of previous theoretical studies where chemical effects on ordering should be dominant to characterize the SRO tendency.\cite{chem1,chem2}

\begin{figure}[h]
\begin{center}
\includegraphics[width=0.88\linewidth]
{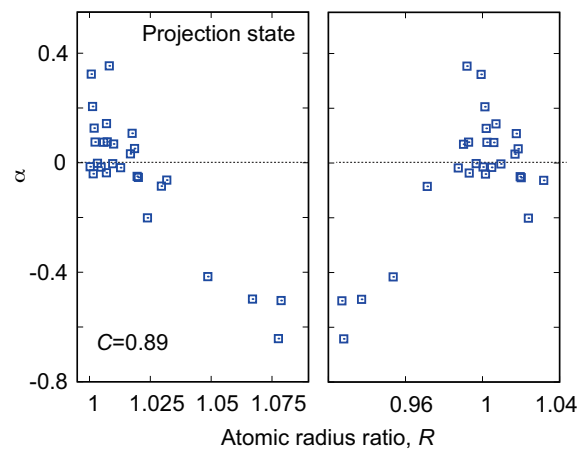}
\caption{Correlation between ratio of effective atomic radius based on projection state and $\alpha$. }
\label{fig:rps}
\end{center}
\end{figure}

In order to further investigate the geometric effects on SRO, we here take another strategy. In term of treating the strain effects on SRO, we should consider changes in volume (or atomic radius) for neighboring, energetically preferred pair between before and after mixing constituents, since actual contribution from changes in volume reflects effective atomic radius in weakly ordering, practical alloys. 
Here, the problem is that SRO depends on the system, it is generally difficult to propose a unified structural parameter to explain difference in $\alpha$, without using explicit information about actual SRO tendency.
This can be practically overcome by using the volume of PS measured from linear average of that for constituents: We have shown that structure of the used PS is interpreted as partially-averaged structure where number of considered unlike-atom pair (or like-atom pair) exceeds their bulk average (i.e., $\left\{ \Braket{q_{1}}_{1}^{\left( r\pm \right)}, \cdots, \Braket{q_{f}}_{1}^{\left( r\pm \right)} \right\}$ as described above). 
This means that volume of the PS directly reflects the changes in volume by mixing constituents to form energetically preferable pairs. Therefore, when only chemical effects contribute to SRO and no geometric (particularly, changes in volume) effects come into play, the atomic ratio for PS measured from volume of constituents should become 1. 
With these considerations, we show in Fig.~\ref{fig:rps} ratio of atomic radius for PS ($R_{\textrm{PS}}$) in terms of linear average of that for constituent elements ($R_{\textrm{ave}}$). Left-hand figure corresponds to that ratio $R$ is define as $R=R_{\textrm{PS}}/R_{\textrm{ave}}$ ($R=R_{\textrm{ave}}/R_{\textrm{PS}}$) when $R_{\textrm{PS}} > R_{\textrm{ave}}$ ($R_{\textrm{ave}} > R_{\textrm{PS}}$), while right-hand figure as $R=R_{\textrm{PS}}/R_{\textrm{ave}}$.
In contrast to Fig.~\ref{fig:rgd}, Fig.~\ref{fig:rps} exhibits explicit linear correlation between $\alpha$ and atomic radius ratio, $R$. 
Particularly, binary alloys with SRO of clustering tendency ($\alpha > 0$) have $R\le 1.02$. 
Again, since SRO is determined by competition between chemical and geometric effects, it is natural that alloys with ordering tendency can have $R$ close to 1:
From Fig.~\ref{fig:rps}, such alloys all exhibit slightly negative $\alpha$, i.e., they show weak ordering tendency. Alloys with strong ordering tendency with $\alpha \le -0.1$ all have large value of $R\ge 1.02$. 
These indicate that for fcc-based disordered phases, alloys with strong ordering tendency are mainly dominated by geometric effects over chemical effects, to reduce strain energy due to pronounced changes in volume by forming energetically preferable like-atom pair with respect to that in unary system. Meanwhile, alloys with weak ordering tendency originally have smaller changes in volume with mixing constituents, whose ordering is determined by combination of chemical as well as geometric effects. 
In the future study, it is fundamentally of interest whether universal threshold for atomic radius exists, to classify whether a given alloy exhibit ordering or clustering tendency on different lattices. 


\section{Conclusions}
Based on first-principle calculation, short-range order in fcc-based disordered alloys is re-examined in terms of underlying geometric nature of the crystalline solids. We confirm that  convensional Goldschmidt or DFT-based atomic radius for constituents have no significant correlation with SRO, which is consistent with the commonly-believed statement established by previous experimental and/or theoretical studies. Despite this fact, we find that atomic radius for specially selected microscopic structure, derived from information only about underlying geometry, can succesfully classify whether a given alloy exhibit ordering or clustering tendency. These facts strongly indicate that geometric effects, particularly changes in atomic radius due to forming actual, energetically prefereble mixing of constituents from unary system, plays central role that can practically compete with chemical effects to determine short-range order for binary alloys.

\section*{Acknowledgement}
This work was supported by a Grant-in-Aid for Scientific Research (16K06704) from the MEXT of Japan, Research Grant from Hitachi Metals$\cdot$Materials Science Foundation, and Advanced Low Carbon Technology Research and Development Program of the Japan Science and Technology Agency (JST).

\end{document}